\documentclass[10pt]{article}
\usepackage{epsfig}
\usepackage{amssymb}
\usepackage{amsmath}

\usepackage{txfonts}
\usepackage{graphicx}
\usepackage{color}
\usepackage{colortab}
\usepackage{pstricks}
\usepackage{enumerate}
\usepackage[normalem]{ulem}

\usepackage{natbib}
\bibpunct{(}{)}{;}{a}{}{,} 
\definecolor{seda}{gray}{.83}

\newgray{lg}{0.85} 
\newgray{mg}{0.7} 

\begin{document}
\begin{center}
{{\large\bf Equations of state in the~Hartle--Thorne model of neutron stars\\selecting acceptable variants of the~resonant switch model of twin\\HF~QPOs in the~atoll source 4U~1636$-$53}}\\
\vspace{3ex}
 {\large Z.~~S~t~u~c~h~l~\'{\i}~k,~~M.~~U~r~b~a~n~e~c,~~A.~~K~o~t~r~l~o~v~\'{a},\\G.~~T~\"{o}~r~\"{o}~k~~and~~K.~~G~o~l~u~c~h~o~v~\'a}\\
\vspace{3ex}
{Institute of Physics, Faculty of Philosophy and Science, Silesian University in Opava, Bezru\v{c}ovo n\'{a}m. 13,
CZ-74601 Opava, Czech Republic\\
{\small e-mail: zdenek.stuchlik@physics.cz, andrea.kotrlova@fpf.slu.cz}}
\end{center}

\vspace{3ex}
\begin{center}
{ABSTRACT}
\end{center}
\noindent{\small The~Resonant Switch (RS) model of twin high-frequency quasi-periodic oscillations (HF~QPOs) observed in neutron star binary systems, based on switch of the~twin oscillations at a~resonant point,  has been applied to the~atoll source 4U~1636$-$53 under assumption that the~neutron star exterior can be approximated by the~Kerr geometry. Strong restrictions of the~neutron star parameters $M$ (mass) and $a$ (spin) arise due to fitting the~frequency pairs admitted by the~RS~model to the~observed data in the~regions related to the~resonant points. The~most precise variants of the~RS~model are those combining the~relativistic precession frequency relations with their modifications. Here, the~neutron star mass and spin estimates given by the~RS model are confronted with a~variety of equations of state (EoS) governing structure of neutron stars in the~framework of the~Hartle--Thorne theory of rotating neutron stars applied for the~observationally given rotation frequency $f_\mathrm{rot}\sim 580\,\mathrm{Hz}$ (or alternatively $f_\mathrm{rot}\sim 290\,\mathrm{Hz}$) of the~neutron star at 4U~1636$-$53. It is shown that only two variants of the~RS~model based on the~Kerr approximation are compatible with two EoS applied in the~Hartle--Thorne theory for $f_\mathrm{rot}\sim 580\,\mathrm{Hz}$, while no variant of the~RS model is compatible for $f_{\mathrm{rot}} \sim 290\,\mathrm{Hz}$. The~two compatible variants of the~RS~model are those giving the~best fits of the~observational data. However, a~self-consistency test by fitting the~observational data to the~RS model with oscillation frequencies governed by the~Hartle--Thorne geometry described by three spacetime parameters $M,\,a$ and (quadrupole moment) $q$ related by the~two available EoS puts strong restrictions. The~test admits only one variant of the~RS model of twin HF QPOs for the~Hartle--Thorne theory with the~\citet{Gan-etal} EoS predicting the~parameters of the~neutron star $M \sim 2.10\,\mathrm{M}_{\odot}$, $a \sim 0.208$, and $q/a^2 \sim 1.77$.}
\\\\{\small{{\bf Keywords:}{~\textit{Accretion, accretion disks --- Stars: neutron --- X-rays: binaries}}}}


\section{Introduction}

The high-frequency quasi-periodic oscillations (HF~QPOs) in the~Galactic Low Mass X-Ray Binaries (LMXBs) containing neutron (quark) stars are often demonstrated as two simultaneously observed pairs of peaks (twin peaks) in the~Fourier power spectra corresponding to oscillations at the~upper and lower frequencies $(\nu_{\mathrm{U}}, \nu_{\mathrm{L}})$ that substantially change over time (even in one observational sequence). Most of the~twin HF~QPOs in the~so-called atoll sources \citep[][]{Kli:2006:CompStelX-Ray:} have been detected at lower frequencies $600-800\,\mathrm{Hz}$ vs. upper frequencies $900-1200\,\mathrm{Hz}$, demonstrating a~clustering of the~twin HF~QPOs frequency ratio around 3\,:\,2 \citep[][]{Abr-etal:2005:ASTRN:,Bel-etal:2007:MONNR:RossiXTE,Tor-etal:2008:ACTA:DistrKhZ4U1636-53,Tor-Bak-Stu-Cec:2008:ACTA:TwPk4U1636-53,Mon-Zan:2012:MNRAS:,Wang-etal:2013:MNRAS:,Ste:2014:MNRAS:}.\footnote{In the~black hole sources, twin peaks with fixed pair of frequencies at the~ratio 3\,:\,2 are usually observed and can be explained by the~internal non-linear resonance of oscillations with radial and vertical epicyclic frequencies \citep{Ter-Abr-Klu:2005:ASTRA:QPOresmodel}.}

For some atoll neutron star sources the~upper and lower HF~QPO frequencies can be traced along the~whole observed range, but the~probability to detect both QPOs simultaneously increases when the~frequency ratio is close to ratio of small natural numbers, namely 3\,:\,2, 4\,:\,3, 5\,:\,4 \citep[][]{Tor:2009:ASTRA:ReversQPOs}. The~analysis of root-mean-squared-amplitude evolution in the~group of six atoll sources (4U~1636$-$53, 4U~1608$-$52, 4U~0614$+$09, 4U~1728$-$34, 4U~1820$-$30, 4U~1735$-$44) shows that the~upper and lower HF~QPO amplitudes equal each other and alter their dominance while passing rational frequency ratios ($3:2$ or $4:3$) corresponding to the~datapoints clustering \citep{Tor:2009:ASTRA:ReversQPOs}. Such an~``energy switch effect'' can be well explained in the~framework of non-linear resonant orbital models as shown in \citet{Hor-etal:2009:ASTRA:IntResQPOs}. Another interesting phenomenon related to energy of the~twin HF~QPOs has been recently demonstrated in \citet{Muk-Bha:2012:ApJ}. Further, analysis of the~twin peak HF~QPO amplitudes in the~atoll sources (4U~1636$-$53, 4U~1608$-$52, and 4U~1820$-$30, 4U~1735$-$44) indicates a~cut-off at resonant radii corresponding to the~frequency ratios 5\,:\,4 and 4\,:\,3 respectively, implying a~possibility that the~accretion disc inner edge is located at the~innermost resonant radius rather than at the~innermost stable circular geodesic \citep[ISCO,][]{Stu-Kot-Tor:2011:ASTRA:ResRadKep}. The~situation is different for some of the~Z-sources where the~twin peak frequency ratios are clustered close to 2\,:\,1, and 3\,:\,1 ratios as demonstrated in the~case of Circinus X-1 \citep[][]{Bou-etal:2006:ASTRJ2:}. Then the~resonant radii could be expected at larger distance from the~ISCO than in the~atoll sources \citep{Tor-etal:2010:ASTRJ2:MassConstraints}.

The evolution of the~lower and upper twin HF~QPOs frequencies in the~atoll and Z sources suggests a~rough agreement of the~data distribution with the~so-called hot spot models of HF~QPOs, e.g., the~Relativistic Precession (RP) model prescribing the~evolution of the~upper frequency by the~Keplerian frequency $\nu_{\mathrm{U}}=\nu_\mathrm{K}$ and the~lower frequency by the~precession frequency  $\nu_{\mathrm{L}}=\nu_\mathrm{K} - \nu_r$ \citep{Ste-Vie:1998:ASTRJ2L:,Ste-Vie:1999:PHYRL:} governed by the~radial epicyclic frequency of geodetical circular motion. In rough agreement with the~data are other models based on the~assumption of the~oscillatory motion of hot spots, or accretion disc oscillations, with oscillatory frequencies given by the~geodetical orbital and epicyclic motion. They include the~modified RP1 model \citep{Bur:2005:RAGtime6and7:CrossRef}, the~Total Precession (TP) model \citep{Stu-Tor-Bak:2007:arXiv:}, the~Tidal Disruption (TD) model \citep{Kos-etal:2009:tidal:}, or the~Warp Disc oscillations (WD) model \citep{Kato:2008:b:PASJ:}. In all of them the~frequency difference $\nu_{\mathrm{U}} - \nu_{\mathrm{L}}$ decreases with increasing magnitude of the~lower and upper frequencies, in accord with the~observational data \citep{Bel-etal:2007:MONNR:RossiXTE,Bar-Oli-Mil:2005:MONNR:,Tor-etal:2012:ApJ:}. This property of the~observational data excludes the~epicyclic oscillations model assuming $\nu_{\mathrm{U}} = \nu_{\theta}$ and $\nu_{\mathrm{L}} = \nu_r$ \citep{Urb-etal:2010:ASTRA:DiscOscNS32} that works well in the~case of HF~QPOs in black hole LMXBs \citep{Ter-Abr-Klu:2005:ASTRA:QPOresmodel}.\footnote{Note that quite recently a~special frequency set of HF~QPOs has been reported for the~neutron star binary system XTE~J1701$-$407 that is one of the~least luminous atoll sources with $L_{X} \sim 0.01\,L_{\mathrm{Edd}}$ \citep{Paw-etal:2013:MNRAS:XTEJ1701-407:}. This frequency set resembles observations of the~HF~QPOs in the~microquasars, i.e., black hole binary systems, and it can be explained by the~model of string loop oscillations \citep{Stu-Kol:2015:GenRelGrav:} that works quite well also in the~case of Galactic microquasars GRS~1915$+$105, XTE~1550$-$564, GRO~1655$-$40 \citep{Stu-Kol:2014:PHYSR4:}.}

The $\nu_{\mathrm{U}} / \nu_{\mathrm{L}}$ frequency relations, given by the~models mentioned above, can be compared to the~observational data found for neutron star LMXBs, e.g., data of the~atoll source 4U~1636$-$53 \citep{Bar-Oli-Mil:2005:MONNR:,Tor-etal:2008:ACTA:DistrKhZ4U1636-53,Tor-Bak-Stu-Cec:2008:ACTA:TwPk4U1636-53}, or the~Z-source Circinus X-1 \citep{Bou-etal:2006:ASTRJ2:}. The~parameters of the~neutron star spacetime can be then determined due to the~fits of the~data to the~frequency-relation models. The~rotating neutron stars are described properly by the~Hartle--Thorne geometry \citep{Har-Tho:1968:ASTRJ2:SlowRotRelStarII} characterized by three parameters: mass $M$, internal angular momentum $J$ and quadrupole moment $Q$, or by dimensionless parameters $a = J/M^2$ (spin) and $q = QM/J^2$. In the~special case when $q/a^2 = 1$, the~Hartle--Thorne external geometry reduces to the~well known Kerr geometry if it is expanded up to the~second order in $a$. The~Kerr approximation is very convenient for calculations in strong gravity regime because of simplicity of relevant formulae. It has been shown recently that near-maximum-mass neutron (quark) star Hartle--Thorne models, constructed for any given equation of state (EoS), imply $q/a^2 \sim 1$ and the~Kerr geometry is applicable with high precision in such situations instead of the~Hartle--Thorne geometry \citep{Urb-Mil-Stu:2013:MONNR:,Tor-etal:2010:ASTRJ2:MassConstraints}. Such high-mass neutron stars can be expected at the~LMXB systems due to the~mass increase caused by the~accretion process.

Assuming the~geodesic orbital and epicyclic frequencies related to the~Kerr geometry, the~fitting procedure applied to the~RP model of the~frequency-relation evolution implies mass--spin relation $M(a) = M_{0}\left[1 + k(a + a^2)\right]$ rather than concrete values of the~neutron star parameters $M$ and $a$; for the~Z-source Circinus X-1 there is $M_{0} \sim 2.2\,\mathrm{M}_{\odot}$ and $k \sim 0.5$ \citep{Tor-etal:2010:ASTRJ2:MassConstraints}. The~same mass--spin relations, but with different values of the~Schwarzschild (no-rotation) mass $M_{0} \sim 1.8\,\mathrm{M}_{\odot}$ and the~constant $k \sim 0.75$, were obtained for the~atoll source 4U~1636$-$53 \citep{Tor-etal:2012:ApJ:}. In the~case of the~models similar to the~RP model (RP1, TP), the~same $M(a)$ relations were found, while for the~models TD and WD, the~relations are different -- for details see \citet{Tor-etal:2012:ApJ:}. Quality of the~fits to the~data obtained for individual models is very poor for the~atoll source 4U~1636$-$53. This fact is extensively discussed in \citet{Tor-etal:2012:ApJ:}. Bad fitting of observational data with the~frequency-relation models was found also in \citet{Lin-etal:2011:ApJ:} for the~atoll source 4U~1636$-$53 and the~Z-source Sco X-1 for some models of the~HF~QPOs with the~frequency relations given by some phenomena of non-geodesic origin \citep{Mil-Lam-Psa:1998:ApJ:,Zha-etal:2006:MONNR:kHzQPOFrCorr,Muk:2009:ApJ:,Shi:2011:}.

The bad fitting of the~data distribution in the~atoll sources by the~frequency-relation models of HF~QPOs based on the~assumption of the~geodesic character of the~oscillatory frequencies invoked attempts to find a~correction of a~non-geodesic origin reflecting some important physical ingredients, as influence of the~magnetic field of the~neutron star onto slightly charged innermost parts of the~disc \citep{Bak-etal:2010:CLAQG:MagIndNonGeo,Kov-Stu-Kar:2008:CLAQG:OffEqOrb}, of thickness of non-slender oscillating tori \citep{Str-Sra:2009:CLAQG:EpiOscNonSleKerrBH}, or of oscillating string loop model \citep{Stu-Kol:2012:JCAP:,Stu-Kol:2014:PHYSR4:,Cre-Stu:2013:PRE}. Such modifications of the~frequency-relation models could make the~fitting procedure better as shown for a~simple toy model in \citet{Tor-etal:2012:ApJ:}. However, in all these cases, some additional free parameter has to be introduced along with the~spacetime parameters of the~neutron star. Some relevant modifications can be also obtained in the~framework of models related to the~braneworld compact objects \citep{Stu-Kot:2009:GENRG2:OrResDiBraKBH,Sch-Stu:2009:GENRG2:ProEmRingBran}.

Therefore, the~Resonant Switch (RS) model of twin peak HF~QPOs has been recently proposed modifying the~standard orbital frequency-relation models in a~way that allows to keep the~assumption of the~relevant frequencies being combinations of the~geodesic orbital and epicyclic frequencies. No non-geodesic corrections are necessary in the~RS~model, although these are not excluded \citep{Stu-Kot-Tor:2012:ACTA:RSmodel,Stu-Kot-Tor:2013:ASTRA:MultiRez} -- the~RS~model considers only the~spacetime parameters of the~neutron star exterior as free parameters. The~RS~model has been applied in the~case of the~atoll source 4U~1636$-$53 \citep{Stu-Kot-Tor:2012:ACTA:RSmodel} and tested for this atoll source by fitting the~observational data using the~frequency relations predicted by the~RS~model as acceptable due to the~neutron star structure theory \citep{Stu-Kot-Tor-Gol:2014:AcA:}.

The fitting procedure predicts the~mass and spin parameters of the~4U~1636$-$53 neutron star with relatively high precision \citep{Stu-Kot-Tor-Gol:2014:AcA:}. Here we test the~frequency relation pairs of the~RS~model giving the~best fits for the~corresponding values of the~mass $M$ and spin $a$ of the~neutron star, using variety of equations of state considered recently in modeling the~rotating neutron stars in the~framework of the~Hartle--Thorne theory. Strong limits implied by the~Hartle--Thorne models can be obtained due to the~precise knowledge of the~rotation velocity of the~4U~1636$-$53 neutron star \citep{Str-Mar:2002:ApJ:}. We are then able to put strong restrictions on validity of the~acceptable frequency-relation variants of the~RS~model.

\section{Resonant switch model of twin HF~QPOs\\in the~4U~1636$-$53 atoll source}

\subsection{The RS~model}

We briefly summarize the~basic ideas of the~RS~model -- for details see \citet{Stu-Kot-Tor:2012:ACTA:RSmodel,Stu-Kot-Tor:2013:ASTRA:MultiRez}. According to the~RS~model a~switch of twin oscillatory modes creating sequences of the~lower and upper HF~QPOs occurs at a~resonant point. Non-linear resonant phenomena are able to excite a~new oscillatory mode (or two new oscillatory modes) and damp one of the~previously acting modes (or both the~previous modes).\footnote{Note that the~switch could occur for other reasons, e.g., due to the~phenomena related to the~magnetic field of neutron stars \citep{Zha-etal:2006:MONNR:kHzQPOFrCorr}.} Switching from one pair of the~oscillatory modes to some other pair will be relevant up to the~following resonant point where the~sequence of twin HF~QPOs ends.

Here, two resonant points at the~disc radii $x_{\mathrm{out}}$ and $x_{\mathrm{in}}$ are assumed ($x = r/(\mathrm{G}M/\mathrm{c}^2)$ is the~dimensionless radius, expressed in terms of the~gravitational radius), with observed frequencies $\nu_{\mathrm{U}}^{\mathrm{out}}$, $\nu_{\mathrm{L}}^{\mathrm{out}}$ and $\nu_{\mathrm{U}}^{\mathrm{in}}$, $\nu_{\mathrm{L}}^{\mathrm{in}}$, being in commensurable ratios
$p^{\mathrm{out}} = n^{\mathrm{out}}:m^{\mathrm{out}}$ and $p^{\mathrm{in}} = n^{\mathrm{in}}:m^{\mathrm{in}}$. Observations of the~twin HF~QPOs in the~atoll systems put the~restrictions $\nu_{\mathrm{U}}^{\mathrm{in}} > \nu_{\mathrm{U}}^{\mathrm{out}}$ and $p^{\mathrm{in}} < p^{\mathrm{out}}$ \citep{Tor:2009:ASTRA:ReversQPOs}. In the~region related to the~resonant point at $x_{\mathrm{out}}$, the~twin oscillatory modes with the~upper (lower) frequency are determined by the~functions $\nu_{\mathrm{U}}^{\mathrm{out}}(x;M,a)$ ($\nu_{\mathrm{L}}^{\mathrm{out}}(x;M,a)$). In the~region related to the~inner resonant point at $x_{\mathrm{in}}$ different oscillatory modes given by the~frequency functions $\nu_{\mathrm{U}}^{\mathrm{in}}(x;M,a)$ and $\nu_{\mathrm{L}}^{\mathrm{in}}(x;M,a)$ occur. All the~frequency functions are assumed to be combinations of the~orbital and epicyclic frequencies of the~geodesic circular motion in the~Kerr backgrounds. Such a~simplification is correct with high accuracy for neutron (quark) stars with large masses, close to maximum allowed for a~given equation of state \citep{Tor-etal:2010:ASTRJ2:MassConstraints,Urb-Mil-Stu:2013:MONNR:}, that can be assumed in the~known atoll or Z-sources because of mass increasing due to the~accretion. Of course, for neutron stars having mass significantly lower than the~maximal allowed mass, the~Hartle--Thorne external geometry reflecting also the~role of the~quadrupole moments of the~neutron star has to be taken into account \citep{Urb-Mil-Stu:2013:MONNR:,Gon-etal:2014:PRD:}.

In the~Kerr spacetime, the~vertical epicyclic frequency $\nu_{\theta}$ and the~radial epicyclic frequency $\nu_{r}$ take the~form  \citep[e.g.,][]{Per-etal:1997:ASTRJ2:,Ste-Vie:1998:ASTRJ2L:,Ter-Stu:2005:ASTRA:}
\begin{equation}
\label{frequencies}
\nu_{\theta}^2 = \alpha_\theta\,\nu_\mathrm{K}^2\,,
\quad
\nu_{r}^2 = \alpha_r\,\nu_\mathrm{K}^2\,,
\end{equation}
where the~Keplerian (orbital) frequency $\nu_\mathrm{K}$ and the~dimensionless quantities determining the~epicyclic
frequencies are given by the~formulae
\begin{eqnarray}
\nu_{\mathrm{K}}&=&\frac{1}{2\pi}\left(\frac{\mathrm{G}M}{r_\mathrm{G}^{~3}}\right)^{1/2}
\left(x^{3/2} + a \right)^{-1} =
\frac{1}{2\pi}\left(\frac{\mathrm{c}^3}{\mathrm{G}M}\right)
\left(x^{3/2}+a\right)^{-1},\\
\alpha_\theta&=& 1-\frac{4a}{x^{3/2}}+\frac{3a^2}{x^{2}}\,,\\
\alpha_r&=&1-\frac{6}{x}+ \frac{8a}{x^{3/2}} - \frac{3a^2}{x^{2}}\,.
\end{eqnarray}
Details of the~properties of the~orbital and epicyclic frequencies can be found in \citet{Ter-Stu:2005:ASTRA:,Stu-Sch:2012:CQG:ObsPhenKerrSSp:}. We can see that any linear combination of the~orbital and epicyclic frequencies depends equally on the~mass parameter $M$, therefore, their frequency ratio becomes independent of $M$. Then the~conditions
\begin{eqnarray}
      &&\nu_{\mathrm{U}}^{\mathrm{out}}(x;M,a) : \nu_{\mathrm{L}}^{\mathrm{out}}(x;M,a) = p^{\mathrm{out}} ,\nonumber\\
      &&\nu_{\mathrm{U}}^{\mathrm{in}}(x;M,a) : \nu_{\mathrm{L}}^{\mathrm{in}}(x;M,a) = p^{\mathrm{in}}
\end{eqnarray}
imply relations for the~spin $a$ in terms of the~dimensionless radius $x$ and the~resonant frequency ratio $p$ that can be expressed as $a^{\mathrm{out}}_{p}(x)$ and $a^{\mathrm{in}}_{p}(x)$, or in an~inverse form $x^{\mathrm{out}}_{p}(a)$ and $x^{\mathrm{in}}_{p}(a)$.

The frequency-relation functions have to meet the~resonant frequencies that can be determined by the~energy switch effect \citep{Tor:2009:ASTRA:ReversQPOs,Stu-Kot-Tor:2012:ACTA:RSmodel}. In the~RS model applied here, two resonant points and two pairs of the~frequency functions are assumed. This enables direct determination of the~Kerr background parameters assumed to govern the~exterior geometry of the~neutron (quark) star.  At the~resonant radii the~conditions
\begin{equation}
        \nu^{\mathrm{out}}_{\mathrm{U}} = \nu^{\mathrm{out}}_{\mathrm{U}}(x;M,a)\,, \quad    \nu^{\mathrm{in}}_{\mathrm{U}} = \nu^{\mathrm{in}}_{\mathrm{U}}(x;M,a)
\end{equation}
are satisfied along the~functions $M^{\mathrm{out}}_{p_{\mathrm{out}}}(a)$ and $M^{\mathrm{in}}_{p_{\mathrm{in}}}(a)$. The~parameters of the~neutron (quark) star are then given by the~condition \citep{Stu-Kot-Tor:2012:ACTA:RSmodel}
\begin{equation}
        M^{\mathrm{out}}_{p_{\mathrm{out}}}(a) = M^{\mathrm{in}}_{p_{\mathrm{in}}}(a)\,.     \label{RS}
\end{equation}
This condition predicts $M$ and $a$ with precision implied by the~error occurring in determination of the~resonant frequencies by the~energy switch effect that is rather high for the~observational data obtained at present state of the~observational devices \citep[see][]{Tor:2009:ASTRA:ReversQPOs,Stu-Kot-Tor:2012:ACTA:RSmodel}. However, the~fitting of the~observational data by the~frequency relations predicted by the~RS~model improves substantially the~precision of determination of the~neutron star parameters and, simultaneously, restricts the~versions of the~RS~model that can be considered as realistic.

Starting from the~results obtained in \citet{Stu-Kot-Tor:2012:ACTA:RSmodel}, in the~present paper we consider pairs of the~frequency relations given by the~RP model \citep{Ste-Vie:1998:ASTRJ2L:,Ste-Vie:1999:PHYRL:}, the~TP model \citep{Stu-Tor-Bak:2007:arXiv:}, and their modifications RP1 \citep{Bur:2005:RAGtime6and7:CrossRef}, and TP1, combined also with the~TD model \citep{Kos-etal:2009:tidal:}, and the~WD model \citep{Kato:2008:b:PASJ:}. The~frequency relations are summarized in Table~\ref{table:models}. In the~RS~model applied to the~source 4U~1636$-$53 the~frequency relations are combined, and the~switch of their validity occurs at the~outer resonant point as described in \citet{Stu-Kot-Tor:2012:ACTA:RSmodel}. For each of the~frequency relations under consideration the~frequency resonance functions and the~resonance conditions determining the~resonant radii $x_{n:m}(a)$ are given in \citet{Stu-Kot-Tor:2012:ACTA:RSmodel}.

\renewcommand{\arraystretch}{1.15}
\begin{table}[h]
\caption{{\small Frequency relations corresponding to individual HF~QPO models.\label{table:models}}}
\begin{center}
\begin{tabular}{lll}
\hline
\textbf{Model} & \multicolumn{2}{c}{\textbf{Relations}}\\
\hline
$\mathbf{RP}\phantom{1}$ & $\nu_{\mathrm{L}} =\nu_{\mathrm{K}}-\nu_{r}$ & $\nu_{\mathrm{U}} =\nu_{\mathrm{K}}$ \\
$\mathbf{RP1\phantom{1}}$ &  $\nu_{\mathrm{L}}=\nu_{\mathrm{K}}-\nu_{r}$ &  $\nu_{\mathrm{U}}= \nu_{\theta}$   \\
\hline
$\mathbf{TP\phantom{1}}$ &  $\nu_{\mathrm{L}}=\nu_{\theta}-\nu_{r}$ &  $\nu_{\mathrm{U}}=\nu_{\theta}$ \\
$\mathbf{TP1\phantom{1}}$ &  $\nu_{\mathrm{L}}=\nu_{\theta}-\nu_{r}$ &  $\nu_{\mathrm{U}}=\nu_{\mathrm{K}}$ \\
\hline
$\mathbf{TD}\phantom{1}$ & $\nu_{\mathrm{L}}=\nu_{\mathrm{K}}$ & $\nu_{\mathrm{U}}=\nu_{\mathrm{K}}+\nu_{r}$ \\
\hline
$\mathbf{WD\phantom{1}}$ & $\nu_{\mathrm{L}}=2\left(\nu_{\mathrm{K}}-\nu_{r}\right)$ & $\nu_{\mathrm{U}}=2\nu_{\mathrm{K}}-\nu_{r}$ \\
\hline
  \end{tabular}
\end{center}
\end{table}


\subsection{Application to the~source 4U~1636$-$53}

The mass $M$ and spin $a$ ranges predicted by the~RS~model with resonant frequencies given by the~energy switch effect are very large \citep[see Table 1 in][]{Stu-Kot-Tor:2012:ACTA:RSmodel}. However, the~ranges can be strongly restricted by fitting the~observational data near the~resonant points by the~pairs of the~frequency relations corresponding to the~twin oscillatory modes. We use the~data of twin HF~QPOs in the~4U~1636$-$53 source as presented and studied in \citet{Tor-etal:2012:ApJ:}, analysed in the~original papers by \citet{Bar-Oli-Mil:2005:MONNR:,Bar-Oli-Mil:2005:AstrNachr:} -- in this case it is immediately clear what is the~extension of the~data related to the~resonant points with frequency ratio 3\,:\,2 and 5\,:\,4, respectively. In the~fitting procedure, based on the~formulae related to the~Kerr spacetime, we applied those switched twin frequency relations predicted by the~RS~model that are acceptable due to the~neutron (quark) star structure theory \citep{Stu-Kot-Tor:2012:ACTA:RSmodel,Stu-Kot-Tor-Gol:2014:AcA:}. All the~resulting twin frequency relations considered in our testing are presented in Table~\ref{table:vysledky-fitu} where the~values of the~mass $M$ and spin $a$ of the~neutron star predicted by the~RS~model and the~related fitting procedure presented in \citet{Stu-Kot-Tor-Gol:2014:AcA:} are explicitly given along with the~corresponding errors. In fitting the~observational data, the~standard least-squares ($\chi^2$) method \citep{Pre-etal:2007:NumReCpp:} has been applied. In the~space of the~lower and upper frequencies, $\nu_{\mathrm{L}}$ and $\nu_{\mathrm{U}}$, the~$\chi^2$-test represents the~minimal (squared) distance of the~frequency relation curve $\nu_{\mathrm{U}}\,(\nu_{\mathrm{L}})$ given by a~model of the~twin HF QPOs from the~observed set of data points. There is
\begin{equation}
      \chi^2 \equiv \sum_{n=1}^{m} \Delta_{n}^2 \,, \qquad      \Delta_n = \mathrm{Min} \left(\frac{l_{n,p}}{\sigma_{n,p}}\right)_{p_\infty}^{p_{\mathrm{ISCO}}}
\end{equation}
where $l_{n,p}$ is the~length of a~line between the~centroid values of the~$n$th measured data point $\left[\nu_{\mathrm{L}}(n),\nu_{\mathrm{U}}(n)\right]$ and a~point $\left[\nu_{\mathrm{L}}(p(n)),\nu_{\mathrm{U}}(p(n))\right]$ belonging to the~relevant frequency curve of the~model; the~points are considered down to the~point corresponding to the~ISCO. The~quantity $\sigma_{n,p}$ denotes the~length of the~part of this line located within the~error ellipse around the~data point \citep{Pre-etal:2007:NumReCpp:}.

The $\chi^2$-test has been applied solely for the~RP, TP and TD frequency relations along the~whole range of the~observational data in \citet{Tor-etal:2012:ApJ:}. However, the~results were quite unsatisfactory, giving $\chi^2 \sim 350$ and $\chi^2/\mathrm{dof} \sim 16$. On the~other hand, the~RS~model enables increase of the~fit precision by almost one order, giving in the~best cases $\chi^2 \sim 55$ and $\chi^2/\mathrm{dof} \sim 2.6$ \citep{Stu-Kot-Tor-Gol:2014:AcA:}.\footnote{The fitting procedure has been realized in the~ranges of $M$ and $a$ predicted by the~RS~model with data given by the~energy switch effect, but we convinced ourselves that outside these ranges the~fits are worse than inside of them.}

\renewcommand{\arraystretch}{1.15}
\begin{table}[h]
\caption{{\small The~best fits of the~observational data and the~corresponding spin and mass parameters of the~neutron star located in the~4U~1636$-$53 source, along with related errors in determining spin and mass of the~neutron star due to the~fitting procedure. The~results are taken, as the~most promising ones, from \citet{Stu-Kot-Tor-Gol:2014:AcA:}. The~two cases of the~neutron star parameters that are in agreement with the~Hartle--Thorne model of the~neutron stars are shaded. They are those corresponding to the~best fits of the~observational data of twin HF~QPOs.\label{table:vysledky-fitu}}}
\begin{center}
\begin{tabular}{lccccc}
  \hline
{\textbf{Combination of models}}  & {$\chi^2_\mathrm{min}$} & {$a$} & {$\Delta a$} & {$M$ $[\mathrm{M}_{\odot}]$} & {$\Delta M$ $[\mathrm{M}_{\odot}]$}\\
\hline
\LCC \lg & \lg & \lg & \lg & \lg & \lg\\
 RP1(3:2) + RP(5:4)  & $55$ & $0.27$ & $0.02$ & $2.20$ & $0.04$ \\
\ECC
 TP(3:2) + RP(5:4)   & $55$ & $0.52$ & $0.02$ & $2.87$ & $0.06$ \\
\LCC \lg & \lg & \lg & \lg & \lg & \lg\\
 RP1(3:2) + TP1(5:4) & $61$ & $0.20$ & $0.01$ & $2.12$ & $0.03$ \\
\ECC
 RP1(3:2) + TP(5:4)  & $62$ & $0.45$ & $0.03$ & $2.46$ & $0.06$ \\
 TP(3:2) + TP1(5:4)  & $68$ & $0.31$ & $0.02$ & $2.39$ & $0.05$ \\
 RP(3:2) + TP1(5:4)  & $72$ & $0.46$ & $0.03$ & $2.81$ & $0.09$ \\
 WD(3:2) + TD(5:4)   & $113$ & $0.34$ & $0.08$ & $2.84$ & $0.21$ \\
\hline
  \end{tabular}
\end{center}
\end{table}


The mass $M$ and spin $a$ ranges determined by the~fitting procedure in \citet{Stu-Kot-Tor-Gol:2014:AcA:} for acceptable combinations of frequency-relation pairs are illustrated in Figures~\ref{graf-M-a-290} and \ref{graf-M-a-580} and summarized in Table~\ref{table:vysledky-fitu}. The~best fit is obtained for the~combination of frequency pairs RP1+RP.

The results of the~fitting procedure will be further tested by confrontation with detailed Hartle--Thorne theoretical models \citep{Har-Tho:1968:ASTRJ2:SlowRotRelStarII,Cha-Mil:1974:MONNR:,Mil:1977:MONNR:} describing slowly rotating neutron stars that are constructed under the~observationally given constraint of the~rotation frequency $580\,\mathrm{Hz}$ (or $290\,\mathrm{Hz}$)  relevant for the~4U~1636$-$53 neutron star \citep{Str-Mar:2002:ApJ:}, using the~variety of widely accepted equations of state that were studied in \citep{Urb-Mil-Stu:2013:MONNR:}. We assume a~detailed test of a~much more extended  family of acceptable equations of state in a~future paper.

The results of the~RS~model have to be related in future to the~limits on the~4U 1636$-$53 neutron star parameters indicated by other possible observational phenomena. In fact, a~preliminary result of simultaneous treatment of the~twin peak HF~QPOs and profiled (X-ray) spectral lines indicates the~neutron star mass to be $M \sim 2.4\,\mathrm{M}_{\odot}$ \citep{Sanna-etal-poster-IAU-Peking-2012} that gives an~important restriction on the~results of the~RS~model and restricts substantially the~variety of allowed combinations of frequency relations used in the~RS~model. However, we clearly need more detailed study of the~profiled spectral lines based on the~precise predictions of the~character of the~external spacetime of the~neutron star.

\section{Hartle--Thorne model of rotating neutron stars}

The Hartle--Thorne theory represents a~standard approximative method of constructing models of compact stars (neutron stars, quark stars, white dwarfs) within general relativity, assuming rigid and slow rotation of the~stars \citep{Har:1967:ASTRJ2:,Har-Tho:1968:ASTRJ2:SlowRotRelStarII}. It is treating deviations away from the~spherical symmetry as perturbations with terms up to a~specified order of the~rotational angular velocity $\Omega$ of the~compact star. Going up to second order in $\Omega$, the~theory gives the~lowest order expressions for the~frame dragging $\omega$, the~moment of inertia $I=J/\Omega$, with $J$ denoting the~angular momentum of the~star, the~shape distortion caused by centrifugal effects, the~quadrupole moment $Q$ and the~change in the~gravitational mass due to rotation $\delta M$. Recent results indicate that the~slow rotation approach is quite correct for all the~observed rotating neutron stars, even in the~case of the~fastest observed pulsar PSR~J1748$-$2446ad with rotational frequency $\sim 716\,\mathrm{Hz}$ \citep{Urb-Mil-Stu:2013:MONNR:}. For very fast rotation only, near to that giving centrifugal break-up, we have to solve numerically the~full set of the~Einstein equations rather than using the~approximative approach of Hartle--Thorne theory -- see models presented in \citet{Bon-Gou-Mar:1998:PHYSR4:} and \citet{Ste:2003:LivRevRel:}.

For our purposes, the~second-order slow-rotation Hartle--Thorne approximative theory developed in \citet{Har-Tho:1968:ASTRJ2:SlowRotRelStarII} and \citet{Cha-Mil:1974:MONNR:} is quite appropriate because of the~rotation frequency observed for the~neutron star in 4U~1636$-$53. Then the~Hartle--Thorne geometry describing both the~internal and external spacetime takes in the~geometric units with $\mathrm{c}=\mathrm{G}=1$ the~form
\begin{eqnarray}
\mathrm d s^2&=&-\mathrm{e}^{2\nu_0}\left[1+2h_0(r)+2h_2(r)P_{2}(\theta)\right] \mathrm dt^2 \nonumber \\\
&&+\mathrm{e}^{2\lambda_0}\left\{1+\frac{\mathrm{e}^{2\lambda_0}}{r}\left[2m_0(r)+2m_2(r)P_{2}(\theta)\right]
\right\}\mathrm d r^2 \nonumber \\
&& + r^2\left[1+2k_2(r)P_{2}(\theta)\right]\left\{\mathrm d \theta^2 + \left[\mathrm d \phi-\omega(r)\mathrm d t\right]^2\sin^2 \theta \right\} ,
\label{eq:HTMetricInternal}
\end{eqnarray}
where $\nu_0$, $\lambda_0$ and the~coordinates are identical with corresponding spherical non-rotating solution, $\omega(r)$ is a~perturbation of order $\Omega$, representing the~frame dragging, and $h_{0}(r)$, $h_{2}(r)$, $m_{0}(r)$, $m_{2}(r)$, $k_{2}(r)$ are perturbations of order $\Omega^2$. All of these perturbations are functions of the~radial coordinate only. The~non-spherical angular dependence is determined by the~second-order Legendre polynomial $P_{2}(\theta) = \frac{1}{2}\left(3 \cos^2\theta - 1\right)$.

All the~perturbation functions have to be calculated under appropriate boundary conditions at the~centre and at the~surface of the~compact star. The~second-order perturbations are labeled with a~subscript indicating their multipole order: $l=0$ for spherical perturbations, $l=2$ for the~quadrupole perturbations representing the~deviation away from the~spherical symmetry. By matching the~internal and external solution at the~star surface, the~external parameters of the~compact star as measured by distant observers can be calculated: the~mass $M$, angular momentum $J$ and quadrupole moment $Q$ that fully characterize the~external gravitational field in the~slow-rotation approximation, if one is retaining only perturbations up to second order.

To construct the~internal solution, the~Einstein equations are solved with the~source term given by the~energy momentum of a~perfect fluid. Rigid rotation of an~axisymmetric configuration means that the~four velocity has components
\begin{equation}
       U^t = \left[-\left(g_{tt} - 2\Omega g_{t\phi} + \Omega^2 g_{\phi\phi}\right)\right]^{1/2}, \quad U^{\phi} = \Omega U^{t} \, .
\end{equation}
The derivation of the~equations for the~perturbation quantities together with boundary conditions has been given in detail in \citet{Har-Tho:1968:ASTRJ2:SlowRotRelStarII,Cha-Mil:1974:MONNR:,Mil:1977:MONNR:}. We will not repeat this in the~present paper, using the~same procedures as those presented in \citet{Mil:1977:MONNR:}.

\section{Equations of state}

The crucial ingredient of the~compact star models is the~equation of state describing properties of matter constituting them. Neutron stars are expected to consist of neutrons closely packed in $\beta$-equilibrium with protons, electrons and at high densities also muons, hyperons, kaons and possibly other particles. Their central densities correspond to microphysics that is not well understood, therefore, they serve as laboratories of nuclear matter under extreme conditions, giving complementary information to those obtained in the~collider experiments.

A wide range of approaches for nucleon-nucleon interactions and their role in modeling of the~structure of neutron stars has been used -- see the~review by \citet{Lat-Pra:2007:PhysRep:}. An~alternative to the~standard neutron star picture is represented by quark stars consisting partially or fully from deconfined quarks. The~most radical version of this approach is represented by strange stars consisting entirely from deconfined quarks \citep{Far-Jaf:1984:PHYSR4:,Hae-Zdu-Sch:1986:ASTRA:,Col-Mil:1992:ApJ:}. It is based on the~suggestion of \citet{Wit:1984:PHYSR4:} that matter consisting of equal numbers of up, down and strange quarks represent the~absolute ground state of strongly interacting matter. It is important to mention that the~strange stars have to be bound together by a~combination of the~strong and gravitational forces, in contrast to neutron stars where only the~gravity is responsible for the~binding. Here we restrict attention on the~equations of state governing the~neutron stars only.

We consider a~set of neutron-star matter equations of state that are based on various approaches -- following the~recent study of the~neutron star properties related to the~behavior of the~quadrupole moment \citep{Urb-Mil-Stu:2013:MONNR:}. We give a~brief review of these equations of state, details can be found in \citet{Urb-Mil-Stu:2013:MONNR:} and the~original literature.

We choose relatively wide set of Skyrme parameterizations, whose labels are starting with S; see \citet{Sto-etal:2003:PHYSR3:} for details of Skyrme potential and diferences between various parameterizations. We use two variants of APR model based on the~variational theory reflecting the~three body forces and the~relativistic boost corrections \citep{Akm-Pan:1998:}. APR corresponds to  $A18 + \delta v + \mathrm{UIX}^*$, while APR2 corresponds to $A18 + \mathrm{UIX}$, where relativistic boost corrections are not included. The~UBS equation of state is based on the~relativistic Dirac--Brueckner--Hartree--Fock mean field theory \citep{Urb-Bet-Stu:2010:ACTA:ObsTestNSMeanField} and correspond to model originally labeled as H. The~non-relativistic Brueckner--Hartree--Fock theory is represented by the~equation of state labeled as BBB2 \citep{BBB2}. We also use GlendNH3 \citep{GlendNH3}  and BalbN1H1 \citep{BALBN1H1} equation of state including the~hyperons at high densities. FPS is the~very well know EoS and has been used very often in the~past \citep{Lor-etal:1993:} and BPAL12 is very soft EoS giving very low maximum mass \citep{BPAL12}. Stiff equations of state were recently constructed in the~framework of the~auxiliary field diffusion Monte Carlo technique and are labeled as Gandolfi \citep{Gan-etal}. Our selection of EoS represents wide range of possible models, however some of these do not meet current observations \citep{Ste-Lat-Bro,Demorest,Antoniadis}. Here, we focus on the~selection of the~acceptable variants of the~RS~model using the~Hartle--Thorne theory of neutron stars that is properly applied to the~restricted set of the~equations of state.

It is useful to give the~maximal values of the~neutron star parameters $M$ and $a$ obtained in the~framework of different approaches to the~equation of state. It follows from general relativistic restrictions that mass of a~neutron star cannot exceed $M_{\mathrm{max}} \sim 3\,\mathrm{M}_{\odot}$ \citep{Rho-Ruf:1974:PHYRL:}. The~realistic equations of state put limit on the~maximal mass of neutron stars $M_{\mathrm{maxN}} \sim 2.8\,\mathrm{M}_{\odot}$  \citep{Pos-etal:2010:LoveNum:} -- the~extremal maximum $M_{\mathrm{maxN}} \sim 2.8\,\mathrm{M}_{\odot}$ is predicted by the~field theory \citep{Mul-Ser:1996:}. The~limit of $M_{\mathrm{maxN}} \sim 2.5\,\mathrm{M}_{\odot}$ is predicted by the~Dirac--Brueckner--Hartree--Fock approach in some special case \citep{Mut-etal:1987:} or by some Skyrme models. The~variational approaches \citep{Akm-Pan:1997:,Akm-Pan:1998:} and other approaches \citep{Urb-Bet-Stu:2010:ACTA:ObsTestNSMeanField} allow for $M_{\mathrm{maxN}} \sim 2.25\,\mathrm{M}_{\odot}$.

On the~neutron star dimensionless spin the~limit of $a < a_{\mathrm{maxN}}=0.7$ has been recently reported on the~base of numerical, non-approximative methods, independently on the~equation of state \citep{Lo-Lin:2011:ApJ:}. The~Hartle--Thorne theory can be well applied up to the~spin $a \sim 0.4$ \citep{Urb-Mil-Stu:2013:MONNR:}.

In the~Hartle--Thorne models of rotating neutron stars the~spin of the~star is linearly related to its rotation frequency. The~rotation frequency of the~neutron star at the~atoll source 4U~1636$-$53 has been observed at $f_\mathrm{rot}\sim 580\,\mathrm{Hz}$ \citep[or $f_\mathrm{rot}\sim 290\,\mathrm{Hz}$, if we observe doubled radiating structure,][]{Str-Mar:2002:ApJ:}. Such a~rotation frequency is sufficiently low in comparison with the~mass shedding frequency, and the~Hartle--Thorne theory can be applied quite well, predicting spins lower enough than the~maximally allowed spin. The~theory of neutron star structure then implies for a~wide variety of realistic equations of state the~spin in the~range \citep{Ste:2003:LivRevRel:,Urb-Mil-Stu:2013:MONNR:}
\begin{equation}\label{limit-na-spin-z-rotace}
    0.1 < a < 0.4\,.
\end{equation}
Of course, the~upper part of the~allowed spin range corresponds to the~rotation frequency $f_\mathrm{rot}\sim 580\,\mathrm{Hz}$, while the~lower part corresponds to $f_\mathrm{rot}\sim 290\,\mathrm{Hz}$. The~related restriction on the~neutron star (near-extreme) mass reads
\begin{equation}\label{limit-na-M-z-rotace}
    M < 2.8\,\mathrm{M}_{\odot}\,.
\end{equation}
Detailed comments on the~precission of the~Kerr geometry approximating the~Hartle--Thorne geometry in dependence on the~spacetime parameters $a$ and $q/a^2$ can be found in \cite{Stu-Kol:2015:GenRelGrav:} -- for the~HF QPO models the~differences induced by the~Kerr approximation could be smaller than five percent for the~dimensionless spin $a < 0.4$.

\section{Testing the~RS~model by equations of state applied\\in the~Hartle--Thorne model of the~neutron star\\in the~source 4U~1636$-$53}

\subsection{Selection of the~relevant variants of the~RS~model}

\begin{figure}
\includegraphics[width=\hsize]{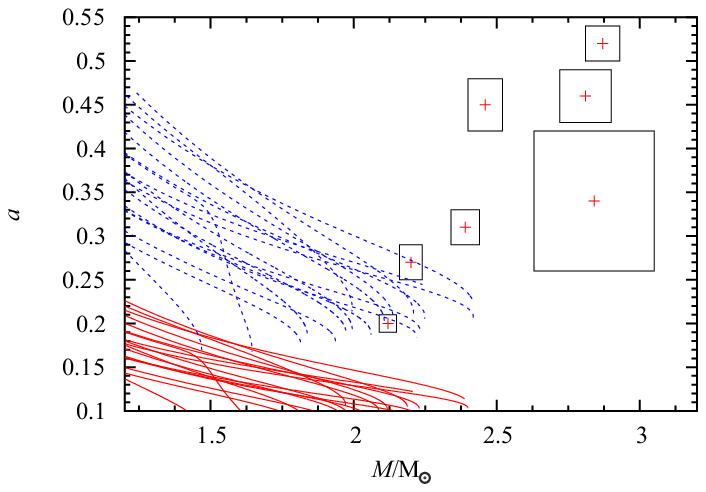}
\caption{Hartle--Thorne models of neutron stars with rotation frequency $f_\mathrm{rot}\sim 290\,\mathrm{Hz}$ (red) and $f_\mathrm{rot}\sim 580\,\mathrm{Hz}$ (blue), for a~variety of EoS considered in \citet{Urb-Mil-Stu:2013:MONNR:}: SkT5, Sk0', Sk0, SLy4, Gs, SkI2, SkI5, SGI, SV -- Skyrme equations \citep{Sto-etal:2003:PHYSR3:}, UBS equation \citep{Urb-Bet-Stu:2010:ACTA:ObsTestNSMeanField}, FPS equation \citep{Lor-etal:1993:}, APR~\citep{Akm-Pan:1998:}, BBB2 equation \citep{BBB2}, BPAL12 equation \citep{BPAL12}, BalbN1H1 equation \citep{BALBN1H1}, GlendNH3 equation \citep{GlendNH3}, APR2 equation \citep{Akm-Pan:1998:}, Gandolfi equation \citep{Gan-etal}. Each equation of state applied in the~Hartle--Thorne model predicts for a~fixed rotational frequency a~sequence of stable states represented by a~curve in the~$M$--$a$ plane; its final point indicates an~instability. Clearly, all predicted values of the~4U~1636$-$53 neutron star spacetime parameters $M$, $a$ related to $f_\mathrm{rot}\sim 290\,\mathrm{Hz}$ are located outside the~predictions of the~Hartle--Thorne models.\label{graf-M-a-290}}
\end{figure}

\begin{figure}
\includegraphics[width=\hsize]{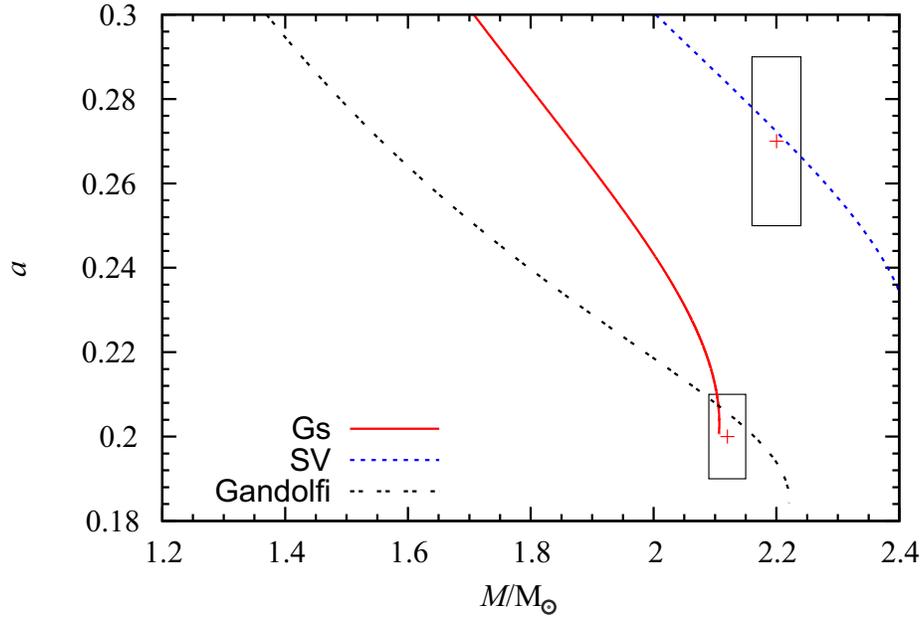}
\caption{Hartle--Thorne models of neutron stars with rotation frequency $f_\mathrm{rot}\sim 580\,\mathrm{Hz}$ are illustrated for the~$a(M)$ dependencies related to the~two best fits of the~observational data by the~RS model. The~acceptable sequences of equilibrium configurations predicted by the~Hartle--Thorne model are constructed for two equations of state (Skyrme SV EoS for RP1+RP variant, and Gandolfi EoS for the~RP1+TP1 variant). Notice that the~variant of the~RS~model predicting the~lower parameters of $M=2.12\,\mathrm{M}_{\odot}$ and $a=0.20$ is also very close to the~final state given by the~sequence of equilibrium states predicted by the~Skyrme Gs EoS, indicating possibility of an~instability of the~neutron star in the~4U~1636$-$53 system in near future.\label{graf-M-a-580}}
\end{figure}

The resulting limits on the~mass $M$ and spin $a$ of the~4U~1636$-$53 neutron star implied by the~data fitting procedure realized in the~framework of the~RS~model of HF~QPOs are presented in Table~\ref{table:vysledky-fitu} taken from \citet{Stu-Kot-Tor-Gol:2014:AcA:} and reflected in Figures~\ref{graf-M-a-290} and \ref{graf-M-a-580}; the~precision of the~mass and spin estimates is also reflected in Figures~\ref{graf-M-a-290} and \ref{graf-M-a-580}. The~Hartle--Thorne models are constructed for a~variety of acceptable equations of state discussed above \citep[and studied in][]{Urb-Mil-Stu:2013:MONNR:} for both possible rotational frequencies of the~4U~1636$-$53 neutron star. We use nine parameterizations of the~Skyrme equation of state (SkT5, Sk0', Sk0, SLy4, Gs, SkI2, SkI5, SGI, SV) and other nine equations of state (UBS, APR, FPS, BBB2, BPAL12, BalbN1H1, GlendNH3, Gandolfi, APR2) that well represent the~variety of eqations of state. The~results of the~Hartle--Thorne model that are calculated for the~equations of state under consideration are illustrated in the~$M$--$a$ plane in Figure~\ref{graf-M-a-290} for both the~assumed neutron star rotation frequencies $f_\mathrm{rot}\sim 290\,\mathrm{Hz}$ and
$f_\mathrm{rot}\sim 580\,\mathrm{Hz}$. Clearly, all the~mass-spin dependencies constructed for $f_\mathrm{rot}\sim 290\,\mathrm{Hz}$ can be excluded. In Figure~\ref{graf-M-a-580} we give detailed picture of fitting the~mass and spin range implied by the~acceptable variants of the~RS model by the~$a(M)$ curves constructed for the~acceptable EoS with the~rotation frequency $f_\mathrm{rot}\sim 580\,\mathrm{Hz}$.

We immediately see that no equation of state allows to construct a~Hartle--Thorne model that can fit the~RS~model data, if we assume the~rotational frequency of the~4U 1636$-$53 neutron star $f_\mathrm{rot}\sim 290\,\mathrm{Hz}$. For the~rotational frequency $f_\mathrm{rot}\sim 580\,\mathrm{Hz}$, the~Hartle--Thorne models give very interesting restrictions that are in significant agreement with results of the~fitting the~HF~QPO data in the~framework of the~RS~model. The~Hartle--Thorne model based on the~Skyrme equation of state SV meets with high precision the~prediction of the~RP1+RP version of the~RS~model that gives the~best fit to the~twin peak HF~QPO data observed in the~4U~1636$-$53 source for the~neutron star parameters $M \sim 2.20\,\mathrm{M}_{\odot}$ and $a \sim 0.27$.\footnote{The same precision of the~fit, namely $\chi^2 \sim 55$, is obtained for the~TP+RP version of the~RS~model. However, in this case the~predicted mass and spin, $M \sim 2.87\,\mathrm{M}_{\odot}$ and $a \sim 0.52$, are outside the~values acceptable by the~neutron star models.} The~Hartle--Thorne model based on the~Gandolfi equation of state meets with high precision the~prediction of the~RP1+TP1 version of the~RS~model that gives the~second best fit to the~observational data of the~twin HF~QPOs in 4U~1636$-$53 for a~neutron star having parameters $M \sim 2.12\,\mathrm{M}_{\odot}$ and $a \sim 0.20$. Notice that the~variant of the~RS~model giving the~second best fit is in accord with another version of the~Skyrme equation of state (Gs) which predicts the~neutron star with parameters $M \sim 2.11\,\mathrm{M}_{\odot}$ and $a \sim 0.20$. Such a~result demonstrates that the~4U~1636$-$53 neutron star could be in a~state very close to an~instability, as the~neutron star mass and spin indicated by the~HF~QPO data fitting procedure can correspond to the~final state of the~evolution of the~neutron stars rotating with the~frequency $f_\mathrm{rot}\sim 580\,\mathrm{Hz}$ and governed by the~Skyrme equation of state Gs -- see Figure~\ref{graf-M-a-580}. Predictions of all the~other variants of the~RS~model are located in the~$M$--$a$ plane at positions that are evidently out of the~scope of all the~equations of state considered in the~present paper -- we can expect that this is true even for all the~variants of the~presently known equations of state.

\subsection{Parameters and the~shape of the~neutron star}

Shape of isobaric surfaces $P=\mathrm{const.}$ and the~shape of the~neutron star surface $P=0$ are given by
\begin{equation}
         r(P=\mathrm{const.},\theta) = r_{0}(P) + \xi_{0}(r_{0}) + \xi_{2}(r_{0})P_{2}(\cos\theta)\,,
\end{equation}
where $r_{0}$ is the~spherical coordinate and functions $\xi_{0}$ and $\xi_{2}$ are given by the~relations \citep{Mil:1977:MONNR:}
\begin{equation}
                \xi_{0}(r) = \frac{r[r-2m(r)]p_{0}}{4\pi r^3 P + m(r)}\,,
\end{equation}
\begin{equation}
                 \xi_{2}(r) = \frac{r[r-2m(r)]p_{2}}{4\pi r^3 P + m(r)}
\end{equation}
and
\begin{equation}
                 p_{2}(r) = -h_{2}(r) - \frac{1}{3}r^2 \mathrm e^{-2\nu_{0}}\omega^{2}\,.
\end{equation}
The calculations have been realized using the~detailed set of equations presented in \citet{Mil:1977:MONNR:}. Then equatorial and polar radii governing the~surface shape of the~rotating neutron star read
\begin{equation}
                 R_{\mathrm{eq}} = R_{0}(P) + \xi_{0}(R_{0}) - \frac{1}{2} \xi_{2}(R_{0})\,,
\end{equation}
\begin{equation}
                 R_{\mathrm{pol}} = R_{0}(P) + \xi_{0}(R_{0}) + \xi_{2}(R_{0})\,.
\end{equation}

The agreement of the~Hartle--Thorne neutron star models based on the~Skyrme and Gandolfi equations of state with two best fits of the~observational data of twin HF~QPOs observed in the~source 4U~1636$-$53 enables to predict in detail properties of the~neutron star in this source. Namely, we are able to find along with the~two known parameters, mass $M$ and spin $a$, also the~radius in the~equatorial plane  $R(\theta=\pi/2)$ and along the~symmetry axis $R(\theta=0)$ and whole the~shape of the~neutron star surface, the~moment of inertia $I$, and quadrupole moment $Q$ and its dimensionless form $q=QM/J^2$. Then we can calculate also the~parameter representing compactness of the~neutron stars in dependence on the~latitude
\begin{equation}
                C_{\theta} = \frac{R(\theta)}{2 M} .
\end{equation}
We can consider the~characteristic values of the~compactness parameter related to the~equatorial plane $\theta=\pi/2$ and the~symmetry axis $\theta=0$. Here we give for simplicity the~compactness parameter related to the~basical, spherically symmetric model that is a~starting point of the~Hartle--Thorne models, given by
\begin{equation}
                              C_0 = \frac{R_0}{2 M_0}.
\end{equation}
The results of the~Hartle--Thorne model calculations for all three equations of state giving acceptable agreement with the~data predicted by the~RS~model are presented in Table~\ref{table:H-T-vysledky}. For the~Skyrme equation of state SV, and the~Gandolfi equation of state, the~neutron star parameters are given for the~mass parameter corresponding to the~mean value of the~data fitting \citep{Stu-Kot-Tor-Gol:2014:AcA:}. In the~case of the~Skyrme equation of state Gs, the~mass parameter of the~neutron star ($M \sim 2.11\,\mathrm{M}_{\odot}$) corresponds to the~maximal value predicted by this equation of state, i.e., it gives the~instability point of the~neutron stars governed by this equation of state. This mass parameter is lower than the~related mean value given by the~data fitting, but it falls into the~allowed range of the~mass parameter.

\renewcommand{\arraystretch}{1.15}
\begin{table}[h]
\caption{{\small Parameters of neutron stars predicted by the~selected equations of state giving Hartle--Thorne models compatible with the~fitting of the~twin HF~QPOs observed in 4U~1636$-$53. The~mass is fixed to the~mean value given by the~fitting procedure with exception of the~Skyrme equation of state Gs where it corresponds to the~final state indicating an~instability. The~spin predicted by the~Hartle--Thorne model fits the~range given by the~error determined by the~data fitting. All models predict very compact neutron stars with low value of the~parameter $q/a^2$.\label{table:H-T-vysledky}}}
\begin{center}
\begin{tabular}{ccccccc}
\hline
{EoS} & $M$ $[\mathrm{M}_{\odot}]$ & $R_{\mathrm{eq}}$ $[\mathrm{km}]$ & $a$ & $q$ & $q/a^2$ & $R_0/(2M_0)$ \\
\hline
 Gandolfi & 2.12 & 10.75 & 0.205 & 0.0723 & 1.71 & 1.72 \\
 Gs & 2.11 & 10.84 & 0.201 & 0.0676 & 1.68 & 1.75 \\
 SV & 2.20 & 13.41 & 0.272 & 0.1940 & 2.60 & 2.07  \\
\hline
  \end{tabular}
\end{center}
\end{table}


\section{Self-consistency test by the~Hartle--Thorne geometry}

Assuming that the~external geometry of the~4U~1636$-$53 neutron star can be approximated by the~Kerr geometry, we have found that the~two most precise variants of the~RS model can be fitted by realistic EoS applied in the~Hartle--Thorne model of slowly rotating neutron stars. For the~RP1+RP variant, the~Skyrme SV EoS predicts $M \sim 2.20\,\mathrm{M}_{\odot}$, $a \sim 0.272$, $q/a^2 \sim 2.60$. For the~RP1+TP1 variant, the~Gandolfi EoS predicts $M \sim 2.12\,\mathrm{M}_{\odot}$, $a \sim 0.205$, $q/a^2 \sim 1.71$; for this RS variant, also the~Skyrme Gs EoS gives acceptable values of $M \sim 2.11\,\mathrm{M}_{\odot}$, $a \sim 0.201$, $q/a^2 \sim 1.68$, however, this estimated mass parameter is at the~maximum allowed mass for the~EoS.

We have to realize now a~self-consistency test of the~Kerr geometry approximation applied in fitting the~observational data. We have to check, if fitting the~observational data by the~$\chi^2$-test using the~orbital and epicyclic frequencies related to the~Hartle--Thorne geometry with parameters $M,\,a,\,q$ governed by the~acceptable EoS gives results comparable or better than the~fitting based on the~assumption of the~Kerr geometry approximation. We realize the~self-consistency test in the~following three steps.

\begin{figure}
\includegraphics[width=\hsize]{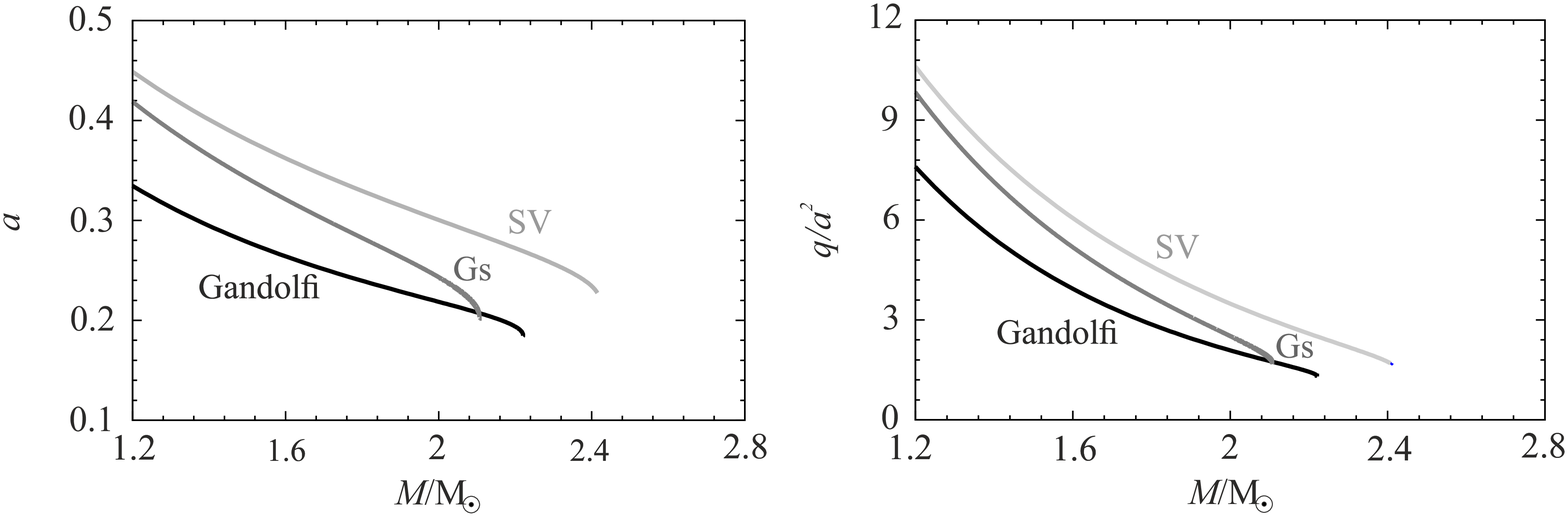}
\caption{Sequence of the~Hartle--Thorne spacetime parameters implied by the~Gandolfi, Skyrme Gs and  Skyrme SV EoS for the~rotational frequency $f_{\mathrm{rot}}=580\,\mathrm{Hz}$. \emph{Left}: Spin ($a$) as a~function of mass ($M$). \emph{Right}: Reduced quadrupole moment ($q/a^{2}$) as a~function of mass ($M$). All sequences are constructed for the~mass interval $1.2\,\mathrm{M}_{\odot} < M < M_{\mathrm{max}}$.\label{EoS}}
\end{figure}

First, we characterize the~sequence of states given by the~Hartle--Thorne model for the~acceptable EoS and the~rotational frequency of the~neutron star $f_{\mathrm{rot}} = 580\,\mathrm{Hz}$ -- see Figure~\ref{EoS}. The~free parameter is the~mass $M$ and we represent the~sequence of the~states by the~functions $a(M)$ and $q/a^2(M)$. For each considered EoS we give the~Hartle--Thorne model for $M>1.2\,\mathrm{M}_{\odot}$, and we follow the~sequence of allowed neutron star states to the~limiting values of the~parameters $M,\,a,\,q/a^2$ corresponding to the~maximal allowed mass for the~considered EoS. For each of the~EoS, the~closest approach of the~Hartle--Thorne model to the~Kerr geometry occurs for the~maximal mass allowed by given EoS. We can see that closest approach to the~Kerr geometry is allowed for the~Hartle--Thorne model based on the~Gandolfi EoS, enabling the~lowest value of $q/a^2 \sim 1.5$ for $M_{\mathrm{max}} \sim 2.24\,\mathrm{M}_{\odot}$. On the~other hand, the~largest difference from the~Kerr geometry approximation are expected for the~Skyrme SV model with the~lowest value of $q/a^2 \sim 1.9$ -- the~maximal mass is highest for this EoS, reaching $M_{\mathrm{max}} \sim 2.4\,\mathrm{M}_{\odot}$.

Second, we define the~orbital and epicyclic radial and vertical frequencies of the~quasicircular geodesic motion in the~Hartle--Thorne geometry, $\nu_{\mathrm{K}}(r;M,a,q)$, $\nu_{r}(r;M,a,q)$, $\nu_{\theta}(r;M,a,q)$ \citep{Abr-etal:2003:ArXiv:,Tor-Bak-Stu-Cec:2008:ACTA:TwPk4U1636-53}. For completeness, we give the~explicit expressions for the~frequencies in the~Appendix.

Third, using the~Hartle--Thorne orbital and epicyclic frequencies, we repeat the~least-square ($\chi^2$-test) fitting procedure for the~same sample of the~observational data in the~4U~1636$-$53 atoll source as those considered in our previous paper \citep{Stu-Kot-Tor-Gol:2014:AcA:}. For the~self-consistency test, we study only the~two selected variants of the~RS model, RP1+RP and RP1+TP1, along the~sequences of the~neutron star parameters $M,\,a,\,q$ related to the~allowed EoS and the~rotational frequency of the~neutron star.\footnote{The fitting of the~data can be done for the~Hartle--Thorne geometry by considering the~neutron star parameters $M,\,a,\,q$ as free parameters. However, such a~fitting is extremely time consuming. We use the~fitting tied to the~EoS and the~rotation frequency of the~neutron star, along the~curve characterized by the~functions $a(M)$, $q(M)$ in the~space of spacetime parameters. This is much faster procedure, being quite relevant for our self-consistency test.}

\begin{figure}
\includegraphics[width=\hsize]{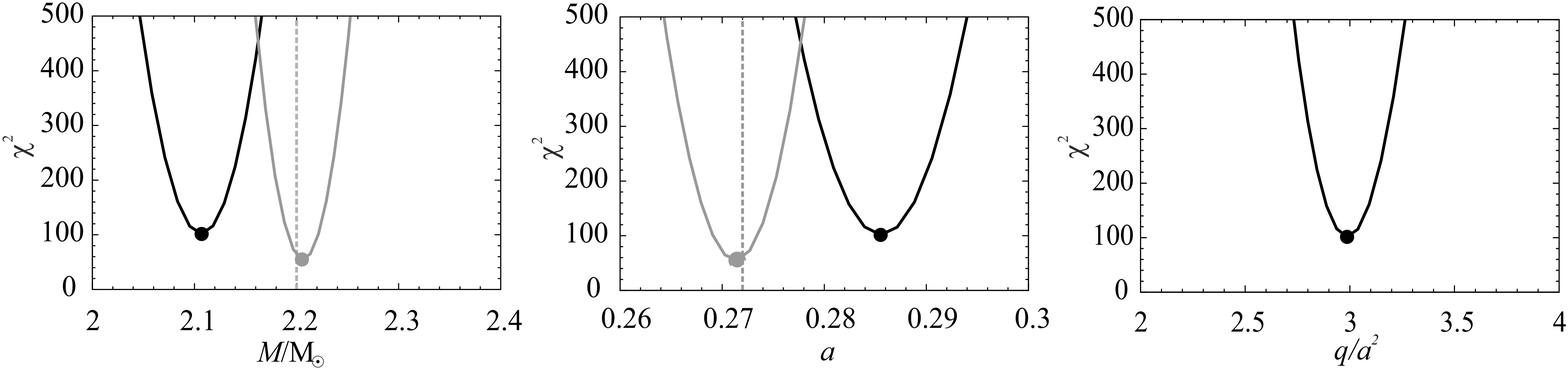}
\caption{The results of the~fitting procedure for the~RP1+RP variant of the~RS model and the~Skyrme SV EoS. $\chi^{2}$ dependency on $M$ (\emph{left}),  $a$ (\emph{middle}), and $q/a^2$ (\emph{right}). The~gray curves correspond to the~Kerr approximation. Dashed lines correspond to values of $M$ and $a$ from Table~\ref{table:H-T-vysledky}. \label{EoS-SV}}
\end{figure}

The results of the~fitting procedure are given in Figure~\ref{EoS-SV} for the~RP1+RP variant of the~RS model and the~Skyrme SV EoS. Along with the~fitting based on the~Hartle--Thorne geometry, we repeat for comparison also the~results obtained under the~Kerr approximation of the~neutron star external spacetime.
The fitting procedure implies the~best fit $\chi^2 = 101$; in comparison to the~best fit based on the~Kerr approximation ($\chi^2=55$, $M \sim 2.20\,\mathrm{M}_{\odot}$, $a \sim 0.272$), the~mass parameter is shifted to lower value of $M \sim 2.11\,\mathrm{M}_{\odot}$, and higher value of spin $a \sim 0.286$. Moreover, at the~values of mass and spin predicted by the~Kerr approximation \citep[when $\chi^2 = 55$,][]{Stu-Kot-Tor-Gol:2014:AcA:}, the~Hartle--Thorne geometry implies $\chi^2 > 1000$. Such a~large discrepancy is caused by relatively large value of the~parameter $q/a^2 \sim 3$ when large errors of the~Kerr approximation are expected. The~resulting value of the~Hartle--Thorne best fit, $\chi^2 = 101$, is too high in comparison with the~Kerr approximation value, $\chi^2 = 55$, and we can conclude that the~RP1+RP variant of the~RS model is not satisfying the~self-consistency test.

\begin{figure}
\includegraphics[width=\hsize]{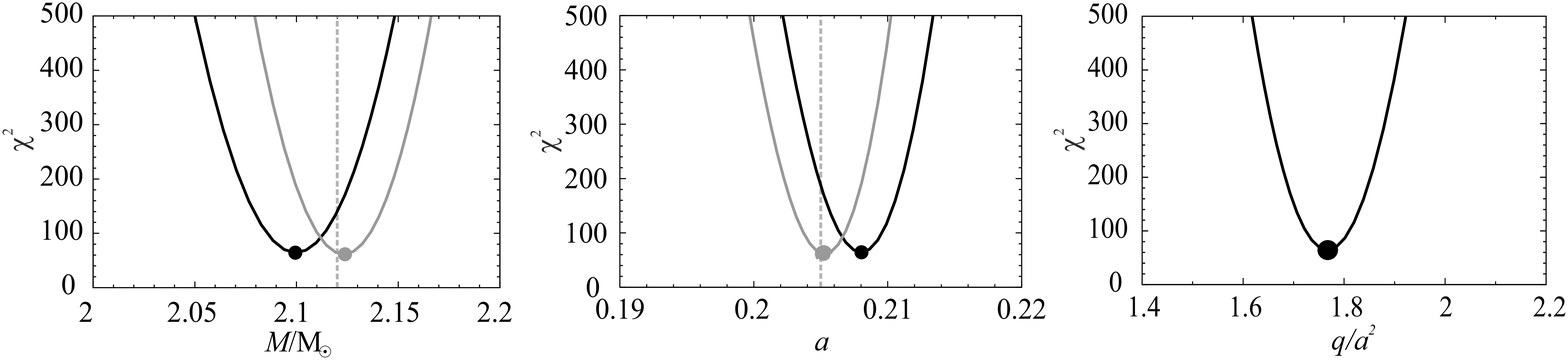}
\caption{The results of the~fitting procedure for the~RP1+TP1 variant of the~RS model and the~Gandolfi EoS. $\chi^{2}$ dependency on $M$ (\emph{left}), $a$ (\emph{middle}) and $q/a^2$ (\emph{right}). The~gray curves correspond to the~Kerr approximation. Dashed lines correspond to values of $M$ and $a$ from Table~\ref{table:H-T-vysledky}.\label{EoS-Gan}}
\end{figure}

For the~RP1+TP1 variant of the~RS model and the~Gandolfi EoS, the~results of the~fitting procedure are presented in Figure~\ref{EoS-Gan}. We again give for comparison the~results of the~best fit obtained for the~Kerr geometry approximation ($\chi^2=61$, $M \sim 2.12\,\mathrm{M}_{\odot}$, $a \sim 0.205$). The~best fit based on the~Hartle--Thorne geometry gives for the~Gandolfi EoS $\chi^2=64$, a~slight decrease of the~mass parameter to $M = 2.10\,\mathrm{M}_{\odot}$ and a~very slight increase of the~spin parameter to $a = 0.208$. Therefore, we can conclude that the~RP1+TP1 model with the~Gandolfi EoS satisfies the~self-consistency test, as both precision of the~fit and the~estimate of the~neutron star parameters are in very good agreement with the~predictions of the~Kerr approximation used in the~fitting procedure. The~precision of the~mass estimate is on the~level of one percent. The~best fit parameter $q/a^2 = 1.77$ is low enough to enable the~high coincidence of predictions of the~Hartle--Thorne geometry and the~Kerr approximation.

\begin{figure}
\includegraphics[width=\hsize]{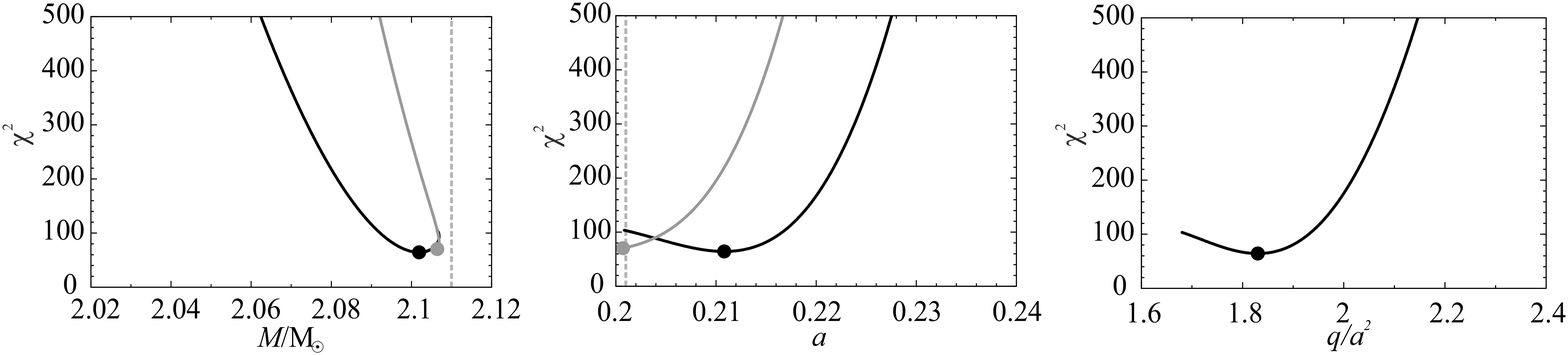}
\caption{The results of the~fitting procedure for the~RP1+TP1 variant of the~RS model and the~Gs EoS.  $\chi^{2}$ dependency on $M$ (\emph{left}), $a$ (\emph{middle}) and $q/a^2$ (\emph{right}).  The~gray curves correspond to the~Kerr approximation. Dashed lines correspond to values of $M$ and $a$ from Table~\ref{table:H-T-vysledky}.\label{EoS-Gs}}
\end{figure}

For the~RP1+TP1 variant of the~RS model and the~Skyrme Gs EoS, the~results of the~fitting procedure are presented in Figure~\ref{EoS-Gs}. We again give for comparison the~results of the~best fit obtained for the~Kerr geometry approximation ($\chi^2=61$, $M \sim 2.12\,\mathrm{M}_{\odot}$, $a \sim 0.201$); however, for this EoS the~maximal allowed mass ($M \sim 2.11\,\mathrm{M}_{\odot}$) is slightly lower than the~Kerr approximation estimate. The~best fit based on the~Hartle--Thorne geometry gives for the~Skyrme Gs EoS $\chi^2=64$, and a~slight decrease of the~mass parameter to $M = 2.10\,\mathrm{M}_{\odot}$, still very close to the~maximal allowed mass for this EOS, and a~slight increase of the~spin parameter to $a = 0.211$. Therefore, we can conclude that the~RP1+TP1 model with the~Skyrme Gs EoS satisfies the~self-consistency test, as both precision of the~fit and the~estimate of the~neutron star parameters are in very good agreement with the~predictions of the~Kerr approximation. The~precision of the~mass estimate is on the~level of one percent again. At the~best fit, the~parameter $q/a^2 = 1.83$ is still low enough to enable the~high coincidence of predictions of the~Hartle--Thorne geometry and the~Kerr approximation. However, the~predicted mass of the~neutron star is very close to the~maximum related to the~Skyrme Gs EoS, bringing some doubts on the~applicability of this EoS for the~source 4U~1636$-$53; if the~Skyrme Gs EoS is the~proper one, than we could expect some strong instability of this source in near future.

The results of the~$\chi^2$-test realized for the~external Hartle--Thorne geometry with parameters governed by the~acceptable EoS are summarized in Table~\ref{HTEoS}. These results give the~self-consistency test of the~results obtained due to the~assumption of Kerr approximation of the~neutron star external spacetime. We can see that only the~RP1+TP1 variant of the~RS model can be considered as surviving the~self-consistency test. Moreover, the~Gandolfi EoS can be considered as the~most plausible one in explaining the~fitting of observational data related to the~twin HF QPOs observed in the~4U~1636$-$53 source. Notice that the~best fits implied by the~Hartle--Thorne geometry give in all considered cases $\chi^2$ value that is higher (worse fit) than in the~case of the~fits based on the~Kerr approximation. Of course, we could obtain better fits by the~Hartle--Thorne geometry for other values of the~neutron star parameters. But in such a~case, the~fits have to be related to the~parameters $M,\,a,\,q$ considered as free parameters, while in our case the~spacetime parameters were confined by the~chosen EoS and the~observed rotation frequency of the~source. Therefore, we cannot exclude that in future an EoS will be discovered that will enable to obtain better fit to the~observational data than those presented in our paper. However, we can note that all EoS considered in our paper give fits worse than those implied by the~Gandolfi EoS.

\renewcommand{\arraystretch}{1.15}
\begin{table}[h]
\caption{{\small Results of the~$\chi^2$-test realized for the~external Hartle--Thorne geometry with parameters governed by the~acceptable EoS.\label{HTEoS}}}
\begin{center}
\begin{tabular}{cccccc}
\hline
{EoS} &   Models & $M$ $[\mathrm{M}_{\odot}]$ &  $a$ & $q/a^2$ &  $\chi^2$  \\
\hline
 Gandolfi & RP1+TP1   &  2.10   &  0.208  &   1.77   & 64 \\
 Gs &  RP1+TP1 &  2.10   &   0.211  &   1.83  &   64 \\
 SV &  RP1+RP  &  2.11   &   0.286  &   2.99  &  101 \\
\hline
  \end{tabular}
\end{center}
\end{table}


\section{Discussion}

The atoll source 4U~1636$-$53 seems to be one of the~best test beds for both the~models of strong gravity phenomena and the~microphysics determining equations of state governing the~internal structure and exterior of neutron stars. This is due to simultaneous availability of relatively good observational data of the~HF QPOs occurring in the~innermost parts of the~accretion disc where the~extremely strong gravity is relevant, enabling thus to put precise restrictions on the~neutron star external spacetime parameters in the~framework of the~RS~model, and the~knowledge of the~rotational frequency of the~neutron star that enables a~precise modeling of the~internal and external structure of rotating neutron stars in the~framework of the~Hartle--Thorne theory, for whole variety of the~equations of state. Strong restrictions on the~acceptable versions of the~RS~model can be obtained, because the~precise knowledge of the~rotational frequency of the~neutron star implies a~narrow evolution line for the~Hartle--Thorne models in the~$M$--$a$ plane that has to be adjusted to relatively precisely determined points in the~$M$--$a$ plane predicted by the~acceptable variants of the~RS~model of the~observed HF~QPOs.

The RS~model can be well tested for the~atoll source 4U~1636$-$53 since this source demonstrates two resonant radii in the~observational data. For all relevant pairs of the~oscillatory frequency relations of the~RS~model, the~range of allowed values of the~mass and dimensionless spin of the~neutron star at 4U~1636$-$53 has been given in \citet{Stu-Kot-Tor:2012:ACTA:RSmodel}. The~most promising frequency pairs predicting the~range of the~4U~1636$-$53 neutron star mass and spin in accord with neutron star structure theory were tested by fitting the~frequency relation pairs to the~observational data on the~HF~QPOs separated into two parts related to the~pair of frequency relations -- only the~frequency relations containing geodesic orbital and epicyclic frequencies, or some combinations of these frequencies, were considered \citep{Stu-Kot-Tor-Gol:2014:AcA:}. Nevertheless, it should be noted that the~cause of the~switch of the~pairs of the~oscillatory modes is not tied to the~resonant phenomena related to the~oscillations governed by the~frequencies of the~geodesic motion. The~switch can be related, e.g., to the~influence of the~magnetic field of the~neutron star and after the~switch the~Alfv\'{e}n wave model can be relevant \citep{Zha-etal:2006:MONNR:kHzQPOFrCorr}. However, limiting the~study to the~resonant phenomena and frequencies of the~geodesic origin, the~number of free parameters of the~model is restricted to the~mass $M$ and dimensionless spin $a$ of the~neutron star, as we are able to demonstrate that the~predicted mass of the~neutron star is large enough to guarantee with high precision independence of the~geodesic frequencies on the~quadrupole moment of the~neutron star and applicability of the~Kerr approximation in decribing the~neutron star external geometry \citep{Urb-Mil-Stu:2013:MONNR:}. Inclusion of the~non-geodesic oscillation modes and non-resonant causes of the~switch is postponed to future studies and could be relevant for some other sources containing neutron stars.

The fitting procedure realized in the~framework of the~RS~model is shown to be more precise by almost one order in comparison to the~standard fitting based on the~individual frequency relations that were used in pairs in the~RS~model \citep{Stu-Kot-Tor-Gol:2014:AcA:}. For example, the~fitting by the~standard RP model predicts the~best fit along the~mass--spin relation $M(a) = M_0 \left[1+0.75(a+a^2)\right]$ with rather poor maximal precision of the~$\chi^2$ test given by $\chi^2 \sim 350$ and $\chi^2/\mathrm{dof} \sim 16$; the~other frequency relations give comparable poor precision \citep{Tor-etal:2012:ApJ:}. Similar results with poor precision were obtained also for models with frequency relations of non-geodesic origin \citep{Lin-etal:2011:ApJ:}. On the~other hand, the~best fit obtained for the~RS~model with frequency relation pair RP1+RP gives $\chi^2 \sim 55$ and $\chi^2/\mathrm{dof} \sim 2.6$ that is quite acceptable due to the~character of the~data distribution \citep{Tor-etal:2012:ApJ:}. The~RP1+TP1 version of the~RS~model predicts the~second best fit with precision that is given by $\chi^2 \sim 61$ and $\chi^2/\mathrm{dof} \sim 2.9$.

Testing the~RS~model by the~Hartle--Thorne theory with fixed rotation frequency and a~variety of equations of state brings another efficient selection of the~variants of the~RS~model. The~results are illustrated by Figures~\ref{graf-M-a-290} and \ref{graf-M-a-580} and clearly demonstrate that only the~two variants of the~RS~model giving the~best results of the~data fitting are acceptable by the~Hartle--Thorne models of the~neutron star structure, if the~rotation frequency $f_\mathrm{rot}\sim 580\,\mathrm{Hz}$. The~RP1+RP version of the~RS~model predicts mean values of mass $M \sim 2.20\,\mathrm{M}_{\odot}$ and spin $a \sim 0.27$ and these are the~data that can be met precisely by the~Hartle--Thorne models -- namely for the~Skyrme equation of state SV. The~RP1+TP1 version of the~RS~model predicts the~mean values of mass $M \sim 2.12\,\mathrm{M}_{\odot}$ and spin $a \sim 0.20$. This mass and spin can be explained by the~Hartle--Thorne model with the~Gandolfi equation of state. It is interesting that the~prediction of the~Hartle--Thorne model based on the~Skyrme equation of state Gs enters the~allowed range of the~mass and spin parameters given by the~RP1+TP1 version of the~RS~model, although it does not reach the~mean value of the~mass parameter, as demonstrated in Figure~\ref{graf-M-a-580}. If this equation of state is the~real one, the~neutron star at the~source 4U~1636$-$53 has to be in a~state very close to instability leading to some form of collapse and dramatic observational phenomena. Mass and spin of the~neutron star predicted by the~other versions of the~RS~model acceptable due to the~data fitting are completely out of the~range of the~$M$--$a$ dependencies predicted by the~Hartle--Thorne model for the~whole variety of available equations of state.

In the~special situations related to accreting neutron stars with near-maximum masses, the~Kerr metric can be well applied in calculating the~orbital and epicyclic geodesic frequencies, as has been done in the~present paper. It should be stressed that the~neutron star mass and spin parameters predicted by the~two relevant findings of frequency pairs are in agreement with the~assumption of near-maximum masses of the~neutron stars -- see Figure~\ref{graf-M-a-580}. For each acceptable equation of state and the~observed rotation frequency of the~neutron star, the~Hartle--Thorne model has been constructed, giving thus not only mass $M$ and spin $a$, but also the~dimensionless quadrupole moment $q$ and other characteristics as the~equatorial radius and the~compactness. The~detailed results of the~Hartle--Thorne model obtained for the~three equations of state that can be in the~play are shown in Table~\ref{table:H-T-vysledky}. The~results clearly demonstrate that in all three cases we obtain a~very compact neutron star, especially for the~Gandolfi and Skyrme Gs equations of state, related to the~second best fit with mass parameter $M \sim 2.12\,\mathrm{M}_{\odot}$ and spin $a=0.2$, having radius $R \sim 3.5\,M$. The~spin exactly predicted by the~Hartle--Thorne model is slightly overcoming the~mean value of the~spin of the~data fitting of the~RS~model, but it belongs to the~allowed range. The~parameter $q/a^2 \sim 1.7$ corresponds to the~external Hartle--Thorne spacetime that is very close to the~Kerr spacetime.

The self-consistency test of the~RP1+TP1 variant of the~RS model by the~Hartle--Thorne geometry, related to the~Gandolfi EoS, confirms this choice, since the~results of the~$\chi^2$-test are very close to those obtained due to the~test by the~Kerr geometry approximation -- both for the~value of $\chi^2$ at the~best fit, and the~close values of the~mass and spin spacetime parameters. Similar results are obtained by the~self-consistency test for the~Hartle--Thorne model related to the~Skyrme Gs EoS. However, the~test confirms also the~conclusion that the~estimated mass has to be very close to the~maximum allowed by the~EoS, lowering thus the~potential relevance of this EoS for the~4U~1636$-$53 neutron star.

For the~RP1+RP variant of the~RS model, the~Hartle--Thorne model using the Skyrme SV EoS, related to the~neutron star with mass $M \sim 2.20\,\mathrm{M}_{\odot}$, the~spin is also predicted with the~high precision $a=0.272$, but the~neutron star is not so extremely compact, having radius $R \sim 4.1\,M$; the~parameter $q/a^2 \sim 2.6$ is too high to approve the~application of the~Kerr geometry in description of the~Hartle--Thorne external spacetime. In fact, the~self-consistency test by the~Hartle--Thorne geometry predicts the~best fit with $\chi^2$ being too high (twice the~estimate due to the~fitting procedure using the~Kerr approximation) to imply relevance of this variant for the~chosen EoS. For this reason the~RP1+RP variant of the~RS model can be considered to be excluded by the~self-consistence test.

\section{Conclusions}

We can conclude that there is a~strong synergy effect of our approach. As expected, the~equations of state applied in the~Hartle--Thorne model of neutron stars fully exclude a~lot of variants of the~RS~model that could be acceptable due to the~fitting procedure to the~HF~QPO data observed in the~4U~1636$-$53 source. Moreover, the~results of the~RS model allow for the~rotation frequency of the~4U~1636$-$53 neutron star $f_{\mathrm{rot}} = 580\,\mathrm{Hz}$, but fully exclude the~possibility of $f_{\mathrm{rot}} = 290\,\mathrm{Hz}$.

The restrictions work effectively in the~opposite direction too -- the~results of the~RS model put significant restrictions on the~relevance of the~equations of state. The~crucial point is that the~self-consistency test by fitting the~observational data by the~RS model with the~orbital and epicyclic frequencies in the~Hartle--Thorne geometry related to the~acceptable EoS excludes one of the~variants of the~RS model predicted by the~fitting under the~Kerr approximation of the~neutron star external geometry, and also the~corresponding EoS. In fact, we have shown that the~RP1+RP variant of the~RS~model along with the~Skyrme SV EoS connected to this variant are excluded by the~self-consistency test giving high value of $\chi^2$ at the~best fit. It is interesting that this happens for the~variant of the~RS~model giving the~best fit to the~data of twin HF~QPOs when the~Kerr approximation of the~oscillatory frequencies has been used.

On the~other hand, the~RP1+TP1 variant of the~RS~model related to a~very stiff Gandolfi EoS goes successfully through the~self-consistency test by the~Hartle--Thorne geometry. In this case, the~resulting best fit gives $\chi^2 = 64$ that is well comparable to the~result obtained for the~Kerr approximation ($\chi^2 = 61$). The~resulting neutron star parameters ($M \sim 2.10\,\mathrm{M}_{\odot}$, $a \sim 0.208$, and $q/a^2 \sim 1.77$) are also very close to those obtained in the~Kerr approximation, demonstrating errors of one percent. Moreover, the~Skyrme Gs EoS used in the~self-consistency test by the~Hartle--Thorne geometry gives also acceptable results, implying a~possibility of the~4U~1636$-$53 neutron star being near the~marginally stable state with mass $M \sim 2.11\,\mathrm{M}_{\odot}$. Of course, the~vicinity of an instability of the~neutron star puts some doubts on the~applicability of the~Skyrme Gs EoS.

We can conclude that in the~framework of the~Hartle--Thorne theory the~EoS imply strong restriction on the~RS model of the~twin HF QPOs observed in the~atoll source 4U~1636$-$53. In fact only the~RP1+TP1 variant of the~RS model satisfies the~self-consistency test. Moreover, it seems that there is only one EoS, namely the~Gandolfi EoS that can be considered as a~fully realistic choice in the~framework of the~modelling the~twin HF QPOs. The~self-consistency test also demonstrates that the~Kerr approximation of the~neutron star external geometry gives very precise estimates for very compact neutron stars having sufficiently low values of the~parameter $q/a^2 < 2$.

It was shown that observations of the~twin HF~QPOs provide tests on equation of state that put limits on the~gravitational mass, and the~spin that is linearly related to the~moment of inertia of the~neutron star. This could provide another test of the~equations of state that allow for existence of neutron stars with $M > 2.0\,\mathrm{M}_\odot$.

For the~equations of state acceptable by the~RS~model we can determine also the~quadrupole moment and  the~shape of the~neutron star surface governed by the~equatorial and polar radii. These quantities have to enter other strong gravity tests of the~4U~1636$-$53 neutron star spacetime parameters predicted by the~twin HF~QPOs, e.g., the~profiled spectral lines generated at the~neutron star surface or at its accretion disc. We believe that such tests could confirm or exclude one of the~two EoS implied by the~acceptable variant of the~RS~model.

Of course, it will be very important to test the~RS model of the~twin HF QPOs and all its consequences for some other neutron star system. We have to check, if the~same variant of the~RS model, and the~same EoS in the~Hartle--Thorne theory of the~neutron stars could be relevant. However, no source similar to the~4U~1636$-$53 neutron star system has been observed. Such sources have to demonstrate sufficiently extended range of the~twin HF QPOs and an indication of two clusters of the~observational data that could be related to different models of twin HF QPOs that could be switched at a~resonant point. Simultaneously, we have to know the~rotation frequency of the~neutron star.

\sloppy
\vspace{3ex}
\noindent
{\small{\bf{Acknowledgements.}}
The~authors acknowledge the~internal grants of the~Silesian University in Opava FPF SGS/11,23/2013. ZS acknowledges the~Albert Einstein Center for Gravitation and Astrophysics supported by the~Czech Science Foundation grant No.~14-37086G. MU and GT acknowledge the~support of the~Czech grant GA\v{C}R 209/12/P740.}


\appendix

\section*{Appendix. Orbital and epicyclic frequencies in Hartle--Thorne geometry}

Circular and epicyclic geodesic motion in the~Hartle--Thorne geometry has been studied in \cite{Abr-etal:2003:ArXiv:,Tor-Bak-Stu-Cec:2008:ACTA:TwPk4U1636-53,Tor-etal:2014:}. Here we only present the~expressions for the~orbital (Keplerian) frequency and the~radial and vertical epicyclic frequencies as given in \cite{Tor-Bak-Stu-Cec:2008:ACTA:TwPk4U1636-53}. Alternative, but equivalent, expressions can be found in \cite{Bos-etal:2014:GraCos:}.

The Keplerian frequency is determined by the~relations
\begin{eqnarray}\label{eq:Kepler}
\nu_{\mathrm{K}}(r;M,a,q) =\frac{c^3}{2\pi GM}\frac{1}{r^{3/2}}\left[1 - \frac{a}{r^{3/2}}+a^2E_{1}(r)+qE_{2}(r)\right],
\end{eqnarray}
where
\begin{eqnarray}
E_{1}(r)&=&[48-80r+4r^2-18r^3+40r^4+10r^5+15r^6-15r^7]\nonumber \\
&&(16(r-2)r^4)^{-1}+\frac{15(r^3-2)}{32}\ln\left(\frac{r}{r-2}\right),\nonumber\\
E_{2}(r)&=&\frac{5(6-8r-2r^2-3r^3+3r^4)}{16(r-2)r}-\frac{15(r^3-2)}{32}\ln\left(\frac{r}{r-2}\right).
\end{eqnarray}
The radial epicyclic frequency $\nu_r$ and the~vertical epicyclic frequency $\nu_{\theta}$ are given by the~relations
\begin{eqnarray}
\nu_{r}^2(r;M,a,q)&=& \left(\frac{c^3}{2\pi GM}\right)^{2}\frac{(r-6)}{r^{4}}[1+
aF_{1}(r)-a^2F_{2}(r)-qF_{3}(r)],
\label{eq:epicrad}\\
\nu_{\theta}^2(r;M,a,q)&=&\left(\frac{c^3}{2\pi GM}\right)^{2}\frac{1}{r^{3}}[1- aG_{1}(r)+a^2G_{2}(r)+qG_{3}(r)],
\label{eq:epicvert}
\end{eqnarray}
where
\begin{eqnarray}
F_{1}(r)&=&\frac{6(r+2)}{r^{3/2}(r-6)},\nonumber\\
F_{2}(r)&=&[8r^4(r-2)(r-6)]^{-1}[384-720r-112r^2-76r^3\nonumber\\
&&-138r^4-130r^5+635r^6-375r^7+60r^8]+A(r),\nonumber\\
F_{3}(r)&=&\frac{5(48+30r+26r^2-127r^3+75r^4-12r^5)}{8r(r-2)(r-6)}-A(r),\nonumber\\
A(r)&=&\frac{15r(r-2)(2+13r-4r^2)}{16(r-6)}\ln\left(\frac{r}{r-2}\right),\nonumber\\
G_{1}(r)&=&\frac{6}{r^{3/2}},\nonumber\\
G_{2}(r)&=&[8r^4(r-2)]^{-1}[48-224r+28r^2+6r^3-170r^4+295r^5\nonumber\\
&&-165r^6+30r^7]-B(r),\nonumber\\
G_{3}(r)&=&\frac{5(6+34r-59r^2+33r^3-6r^4)}{8r(r-2)}+B(r),\nonumber\\
B(r)&=&\frac{15(2r-1)(r-2)^2}{16}\ln\left(\frac{r}{r-2}\right).\nonumber
\end{eqnarray}

\end{document}